\documentclass[a4paper,twocolumn]{autart}

\usepackage{times,bm}

\usepackage{graphicx,psfrag,enumerate,subfigure}
\usepackage{amssymb}
\usepackage{amsmath}
\usepackage{calrsfs}
\usepackage{subeqnarray}
\usepackage{latexsym}

\newtheorem{definition}{Definition}
\newtheorem{theorem}{Theorem}
\newtheorem{remark}{Remark}

\newtheorem{lemma}{Lemma} 
\newtheorem{corollary}{Corollary}

\def\diag{\mathop{\mathrm{diag}}\nolimits} 

\begin{document}
\begin{frontmatter}

\title{Distributed robust estimation over
  randomly switching networks using $H_\infty$ consensus\thanksref{footnoteinfo}}

\thanks[footnoteinfo]{This work was supported by the Australian
  Research Council.}

\author[First]{V.~Ugrinovskii}\ead{v.ugrinovskii@gmail.com} 

\address[First]{School of Engineering and IT, University of NSW
  at the 
  Australian Defence Force Academy, Canberra, ACT, 2600, Australia}

\begin{keyword}                           
Large-scale systems, distributed robust estimation, worst-case transient
consensus, 
vector Lyapunov functions
.    
\end{keyword}

\begin{abstract} 
 The paper considers a distributed robust estimation problem over a network
 with Markovian randomly varying topology. The objective is to deal with
 network variations
 locally, by switching observer gains at affected nodes only. 
 We propose sufficient conditions which guarantee a suboptimal $H_\infty$
 level of relative disagreement of estimates in such observer networks. When the
 status of the network is known globally, these sufficient conditions enable the
 network gains to be computed by solving certain LMIs. When the nodes are
 to rely on a locally available information about the network topology,
 additional rank constraints are used to condition the gains, given this
 information. The results are complemented by necessary conditions which relate
 properties of the interconnection graph Laplacian to the mean-square
 detectability of the 
 plant through measurement and interconnection channels. 
 
\end{abstract}

\end{frontmatter}

\section{Introduction}

One of the motivations for using distributed multisensor networks is to make
the network resilient to loss of communication. This has led to an
extensive research into 
distributed filtering over networks with time-varying,
randomly switching topology. 
In particular, 
the Markovian approach to the analysis and synthesis of 
estimator networks has received a significant attention in relation to the
problems involving random data loss in channels 
with memory which are governed
by a Markov
switching rule~\cite{SS-2009,MMB-2009}.

In addition to capturing memory properties of physical communication channels,
Markovian models allow for other random events in the
network, such as sensor failures and recovery,  to be considered in a
systematic manner within the
Markov jump systems framework. However, the Markov jump
systems theory usually assumes the complete
state of the underlying Markov chain to be known to every  controller or
filter~\cite{LUO1}
. In the context of
distributed estimation and control, this requires each node
of the network to know the complete instantaneous state of the network to
be able to deploy suitable gains. To circumvent such an
unrealistic assumption, the literature focuses on networks whose
communication state is governed by a random process decomposable
into independent two-state Markov processes describing the status of
individual links~\cite{LGM-2006,SS-2009}, even though this typically leads to  design conditions whose complexity
grows exponentially~\cite{LGM-2006}. Also, the
assumption of independence between communication links may not always be
practical, e.g., when dealing with congestions.

The objective of this paper is to develop a distributed filtering technique
which overcomes the need for broadcast of global communication topology and
does not require Markovian segmentation of the network. Our main
contribution is the methodology of robust distributed observer design
which enables the node observers to be implemented in a truly distributed
fashion, by utilizing only locally available information about the system's
connectivity, and without assuming the independence of communication
links. This information structure constraint is a key
distinction of this work, compared with the existing results, e.g.,
\cite{SS-2009,LGM-2006}.  In addition, the proposed methodology allows to
incorporate other random events such as sensor failures and recoveries.

The paper focuses on the case where the plant to be observed, as well
as sensing and communication models are not known perfectly. To deal with uncertain perturbations in the
plant, sensors and communications, we employ the distributed $H_\infty$
filtering framework which has received a significant deal of attention in
the recent literature~\cite{SWH-2010,U6,LaU1}. 
The motivation for
considering $H_\infty$ observers in this paper, instead of Kalman filters~\cite{SS-2009},
is to obtain observers that have guaranteed robustness 
properties. It is well known that the standard Kalman filter 
is sensitive to modelling errors~\cite{PSB}, and
consensus Kalman filters may
potentially suffer from the same shortcomings. 
This explains our interest in robust performance guarantees in the presence of  
uncertainty. 

In contrast to~\cite{SS-2009,SWH-2010}, in this paper the node
estimators are sought to reach relative $H_\infty$
consensus about the estimate of the reference plant. As an extension of the
consensus estimation methodology~\cite{Olfati-Saber-2007}, our approach
responds to the challenge posed by the presence of uncertain perturbations in
the plant, measurements and interconnections. Typically, a perfect consensus
between sensors-agents is not possible due to perturbations. 
To address this challenge, we employ
the approach based on optimization of the transient relative $H_\infty$
consensus performance metric, originally proposed 
in~\cite{U6}. We approach the robust
consensus-based estimation problem from the dissipativity viewpoint, 
using vector storage functions and vector supply rates~\cite{HCN-2004}. 
This allows us to establish both mean-square robust convergence and robust
convergence with probability 1 of the distributed filters under
consideration and guarantee a prespecified level of $H_\infty$
mean-square disagreement between node estimates in the presence of
perturbations and random topology changes.  

The information structure constraint, where the filters must rely on the local
knowledge of the network topology, poses the main challenge in
the derivation of the above-mentioned results. The
standard framework of Markov jump systems is not directly applicable
to the problem of designing locally constrained filters whose information
about the network status is non-Markovian. 
To overcome this difficulty, we adopt the approach recently proposed for 
decentralized control of jump parameter systems~\cite{XiUP1}. It involves a
two-step design procedure. First, an auxiliary distributed estimation
problem is solved under 
simplifying assumption that the complete Markovian network topology is
instantaneously available at each node. However, we seek a solution to
this problem using a network of \emph{non-fragile} estimators subject to
uncertainty~\cite{HC-2000}. Resilience of the auxiliary estimator to
uncertain perturbations is the key 
property to allow this auxiliary 
uncertain estimator network to be modified, at the second step, into an
estimator network which satisfies the information structure constraint and
retains robust performance of the auxiliary design.

An important question in connection with our
distributed observer architecture is concerned
with  requirements on the communication topology under which the 
consensus of node observers is achievable. 
For networks of one- or two-dimensional agents, and
networks consisting of identical agents, conditions for consensus are tightly
related to properties of the graph Laplacian
matrix~\cite{OM-2004,RB-2005,ZLD-2011}. In a more general situation 
involving nonidentical node observers, the role of the interconnection graph
is often hidden behind the design conditions,
e.g., see \cite{SS-2009,SWH-2010}. Our second contribution is to show that
for the distributed 
estimation problem under consideration to have a solution, the standard
requirement for the graph Laplacian to have a simple zero eigenvalue must be
complemented by detectability properties of certain matrix pairs formed by
parameters of the observers and interconnections.  

The paper is organized as follows. The problem formulation is given in
Section~\ref{Distr.Cons}. Section~\ref{sec:design-global} studies an
auxiliary distributed 
estimation problem without the information structure constraints. The
results of this section are then used in Section~\ref{main} where the main
results of the paper are given. Section~\ref{requirements} discusses
requirements on the observer communication
topology. Section~\ref{Example} presents an
illustrating example.

\paragraph*{Notation} $\mathbf{R}^n$ is the real
Euclidean $n$-dimensional vector space, with the norm $\|x\|\triangleq
(x'x)^{1/2}$;  $'$ denotes the transpose of a matrix or a
vector. Also, for a given $P=P'$,
$\|x\|_P=\sqrt{x'Px}$. 
$\mathbf{1}_k\triangleq [1~\ldots~1]'\in \mathbf{R}^k$, and
$I_k$ is the identity matrix in $\mathbf{R}^k$; we will omit the
subscript $k$ when this causes no ambiguity. 
For $X=X'$, $Y=Y'$,
we write $Y>X$ ($Y\ge X$), when  $Y-X$ is positive definite
(positive semidefinite).  $\otimes$
  denotes the Kronecker product of matrices.
$\diag[P_1,\ldots,P_N]$
  is the block-diagonal matrix, whose  diagonal
  blocks are $P_1,\ldots,P_N$. The symbol $\star$ in  position $(k,l)$
of a block-partitioned matrix denotes the transpose of the $(l,k)$ block of
the matrix. $L_2[0,\infty)$ is the Lebesgue space of
$\mathbf{R}^k$-valued vector-functions $z(\cdot)$, defined on $[0,\infty)$,
with the norm $\|z\|_2\triangleq 
\left(\int_0^\infty\|z(t)\|^2dt\right)^{1/2}$.

\section{Problem formulation}\label{Distr.Cons}

\subsection{Networks with Markovian switching topology}\label{networks}

Consider a directed weakly connected
graph $\mathbf{G}=(\mathbf{V},\mathbf{E})$, where $\mathbf{V}=\{1,\ldots,N\}$
is the set of nodes, and $\mathbf{E}\subseteq \mathbf{V}\times
\mathbf{V}$ is the set of edges. The edge $(j,i)$ 
originating at node $j$ and ending at node $i$ represents the event ``$j$
transmits information to $i$''. In accordance with a common convention, we
consider graphs without 
self-loops, i.e., $(i,i)\not\in\mathbf{E}$. However, each node is assumed
to have complete information about its filter, measurements and the
status of incoming communication links. 

We consider two types of random events at each
node. Firstly, node neighborhoods change randomly as a result of random link
dropouts and recovery. Also, to account for sensor adjustments in
response to these changes, as well as sensor failures/recoveries, we 
allow for random variations of the sensing regime at each node. 
Letting $x(t)$, $y_i(t)$ denote an
observed process and its measurement taken at node $i$ at time $t$, and
using a standard linear relation between these quantities
\begin{equation}\label{U6.yi.1.pattern}
y_i=\tilde C_i x+\tilde D_i \xi+{\tilde{\bar
    D}_i}\xi_i, \quad y_i\in\mathbf{R}^r,
\end{equation}
such adjustments are associated with randomly varying coefficients $\tilde C_i$,
$\tilde D_i$, ${\tilde{\bar D}_i}$. These random events are additional
to link dropouts.  This leads us to consider the combined evolution of each
node's neighbourhood and sensing regime.   

\begin{definition}\label{Distinct.N}
For a node $i$, let $\mathbf{V}_i$, $(\tilde C_i,\tilde D_i, \tilde{\bar
  D}_i)$ be its neighbourhood set and the measurement matrix 
triplet, respectively,
at a certain time $t$.  The pair $\{\mathbf{V}_i,(\tilde C_i,\tilde
D_i, \tilde{\bar D}_i)\}$, is said to represent the \emph{local
communication and sensing state} (or simply the \emph{local state}) of node 
$i$ at time $t$. Two states of $i$ at times $t_1$, $t_2$, 
$\{\mathbf{V}_i^1,(\tilde C_i^1,\tilde D_i^1, \tilde{\bar D}^1_i)\}$, 
$\{\mathbf{V}_i^2,(\tilde C_i^2,\tilde D_i^2, \tilde{\bar D}^2_i)\}$
are distinct if\/
$\mathbf{V}_i^1\neq \mathbf{V}_i^2$, or
$(\tilde C_i^1,\tilde D_i^1,
\tilde{\bar D}^1_i)\neq (\tilde C_i^2,\tilde D_i^2, \tilde{\bar
  D}^2_i)$.
\end{definition}
From now on, we associate with every node $i$ the ordered collection of all
its feasible distinct local states and denote the
corresponding index $ \mathcal{I}_i\triangleq \{1,\ldots,M_i\}$.
The time evolution of each local state will be represented by a random mapping
$\eta_i:[0,\infty) \to \mathcal{I}_i$.     

The global configuration and sensing pattern of the network at
any time can be uniquely determined from its local states. This leads us to
define the \emph{global state} of the network as an $N$-tuple
$(k_1,\ldots,k_N)$, where $k_i\in \mathcal{I}_i$. 
Consider the ordered collection of all feasible global states of the
network and let $\mathcal{I}=\{1,\ldots,M\}$ denote its 
index set.  In general, not all
combinations of local states correspond to feasible global states. Owing to
dependencies between network links and/or sensing regimes, the number of
feasible global states may be substantially smaller than the cardinality of
the set $\mathcal{I}_1\times\ldots\times \mathcal{I}_N$ of all
combinations of local states. The one-to-one mapping between
the set of feasible global states  
$\{(k_1,\ldots,k_N)\}$ and its index set $\mathcal{I}$ will be denoted
$\Phi$, i.e., $(k_1,\ldots,k_N)=\Phi(m)$, where $m$ is the index of the
$N$-tuple  $(k_1,\ldots,k_N)$.  Also, we write
$k_i=\Phi_i(m)$, whenever $(k_1, \ldots, k_N)=\Phi(m)$. 

Using the one-to-one mapping $\Phi$, define the \emph{global} process 
$\eta(t)=\Phi^{-1}(\eta_1(t),\ldots,  \eta_N(t))$ to describe the evolution
of the network global state. 
The local state processes $\eta_i(t)$ are related to it as
$\eta_i(t)=\Phi_i(\eta(t))$ $\forall t\ge 0$.  Throughout the paper, we
assume that
$\{\eta (t),t\geq 0\}$ is a stationary Markov random process $[0,\infty)\to
\mathcal{I}$ defined in a
filtered probability space $(\Omega, \mathcal{F}, 
\{\mathcal{F}_t\},\mathsf{P})$, where $\mathcal{F}_t$ denotes a 
right-continuous filtration with 
respect to which $\{\eta(t),t\geq 0\}$ is adapted\footnote{In the sequel,
  we will consider the
  filtration generated by a composite Markov process consisting of
  $\eta$ and error dynamics of the estimator introduced in the next section.}   
  \cite{BC-2009}. 
The $\sigma$-algebra $\mathcal{F}$ is the minimal
$\sigma$-algebra which contains all measurable sets from the filtration
$\{\mathcal{F}_t,t\ge 0\}$. 
The transition probability rate matrix of the Markov chain $\{\eta (t),t\geq 
0\}$ will be denoted $\Lambda=[\lambda_{kl}]_{k,l=1}^M$, with
$\lambda_{kl}\geq 0,k\neq l$ and $\lambda_{kk}=-\sum\limits_{l\neq
  k}{\lambda_{kl}\leq 0,\forall k \in \mathcal{I}}$~\cite{BC-2009}.

Using the global state process $\eta(t)$, the time evolution of the
communication graph can be represented by a random graph-valued process
$\mathbf{G}^{\eta(t)}$, whose value at every time instance is a directed
subgraph of $\mathbf{G}$. It is assumed that for all $t$,
$\mathbf{G}^{\eta(t)}$ is weakly connected and has the same vertex set as
$\mathbf{G}$. When $\eta(t)=k\in\mathcal{I}$, 
$\mathbf{A}^k=[\mathbf{a}_{ij}^k]_{i,j=1,N}$ will denote the adjacency
matrix of the digraph $\mathbf{G}^k=\mathbf{G}^{\eta(t)}$. Note that
$\mathbf{a}_{ij}^k=1$ if and only if $j\in \mathbf{V}_i^{\Phi_i(k)}$. Here
and hereafter, the symbol $\mathbf{V}_i^{k_i}$ describes the neighbourhood
of node $i$ when this node is in local state $k_i$. In accordance with
this notation, $\mathbf{V}_i^{\Phi_i(k)}$ is the neighbourhood
of node $i$ when the network is in global state $k$. Also,  
$p_i^k=\sum_{j=1}^{N}\mathbf{a}_{ij}^k$, 
$q_i^k=\sum_{j=1}^{N}\mathbf{a}_{ji}^k$, and $\mathcal{L}^k$ denote the
in- and out-degrees of node $i$ and the Laplacian matrix of the
corresponding graph $\mathbf{G}^k$, respectively.

We will use the notation $(\eta,\mathcal{G},\Phi)$ to refer to the
switching network described above. Since $\eta(t)$ is 
stationary, then each 
process $\eta_i(t)$ is also stationary. However, in general the local state
processes $\eta_i(t)$ are not Markov, and the components of the
multivariate process $(\eta_1(t),\ldots, \eta_M(t))$ may statistically
depend on each other. Hence our network model allows for
dependencies between links within the network.

\subsection{Distributed estimation with $H_\infty$
  consensus}
Consider a plant described by the equation
\begin{equation}
  \label{eq:plant.1}
  \dot x=Ax+B_2 \xi(t), \quad x(0)=x_0.
\end{equation}
Here $x\in\mathbf{R}^n$ is the state, $\xi(t)\in\mathbf{R}^l$ is a deterministic
disturbance. We assume that $\xi(\cdot)\in
L_2[0,\infty)$, and that the solution of (\ref{eq:plant.1})
exists on any finite interval $[0,T]$, and is $L_2$-integrable on
$[0,T]$. 

Also, consider an observer network $\{\eta,
\mathcal{G},\Phi\}$ whose nodes take measurements of the plant
(\ref{eq:plant.1}) as follows 
\begin{equation}\label{U6.yi.1}
y_i={\tilde C}_i^{\eta_i(t)}x+{\tilde D}_i^{\eta_i(t)}\xi+{\tilde{\bar
    D}_i}^{\eta_i(t)}\xi_i, \quad y_i\in\mathbf{R}^{r_i},
\end{equation}
where $\xi_i(t)\in \mathbf{R}^{l_i}$ represents the deterministic measurement 
uncertainty at sensing node $i$, $\xi_i(\cdot)\in L_2[0,\infty)$. The 
coefficients of equation (\ref{U6.yi.1}) take values in given sets of
constant matrices of compatible dimensions, 
\[
({\tilde C}_i^{\eta_i(t)},{\tilde  D}_i^{\eta_i(t)},
{\tilde{\bar D}_i}^{\eta_i(t)})\in \{(\tilde C_i^k,\tilde D_i^k,\tilde{\bar
  D}_i^k),k\in\mathcal{I}_i\}. 
\]
It will be assumed throughout the paper that $\tilde E_i^k=\tilde D_i^k(\tilde
D_i^k)'+\tilde{\bar D}_i^k(\tilde{\bar D}_i^k)'>0$ for all $i$ and $k\in\mathcal{I}_i$. 

The measurements $y_i$ are processed at node $i$ according to the following
estimation algorithm (cf. \cite{SS-2009,U6,LaU1}): 
 \begin{eqnarray}
    \dot{\hat x}_i&=&A\hat x_i +
    {\tilde L}_i^{\eta_i(t)}(y_i(t)-{\tilde C}_i^{\eta_i(t)}\hat x_i)\nonumber \\
    &+&\sum_{j\in
      \mathbf{V}_i^{\eta_i(t)}}{\tilde K}_{ij}^{\eta_i(t)}(v_{ij}-H_{ij}\hat
    x_i), \quad \hat x_i(0)=0, 
  \label{UP7.C.d.loc} 
\end{eqnarray}
where $v_{ij}$ is the signal received at node $i$ from node $j$,
\begin{eqnarray}
  \label{vij}
  &&v_{ij}=H_{ij}\hat x_j+G_{ij}w_{ij}, \quad v_{ij}\in \mathbf{R}^{r_{ij}},
\end{eqnarray}
$w_{ij}\in \mathbf{R}^{s_{ij}}$ describes the channel uncertainty affecting the
information 
transmission from node $j$ to $i$. It is assumed that $w_{ij}$ belongs  to 
the class of mean-square $L_2$-integrable random disturbances, adapted to the
filtration $\{\mathcal{F}_t,t\ge 0\}$. 

It will be further assumed that
$F_{ij}=G_{ij}G_{ij}'>0$ for all $i$ and $j\in\mathbf{V}_i^{k_i}$,
$k_i\in\mathcal{I}_i$. Also in (\ref{UP7.C.d.loc}), ${\tilde
  L}_i^{\eta_i(\cdot)}$, ${\tilde K}_{ij}^{\eta_i(\cdot)}$ are
matrix-valued functions of the local state process $\eta_i(t)$. These
functions are the design parameters of the algorithm describing innovation
and interconnection gains of the observer (\ref{UP7.C.d.loc}). Note that 
the coupling and
observer gains ${\tilde 
  K}_{ij}^{(\cdot)}$, ${\tilde L}_i^{(\cdot)}$  are required to be
functions of the local state (i.e., functions of $\eta_i$), rather than the
global state. This `locality' information structure 
constraint is additional to the assumption about the Markov nature of the communication
graph; cf.~\cite{SS-2009} where the complete communication
graph was assumed to be known at each node. The problem in this paper is to determine these functions to
satisfy certain robust performance criteria to be presented in
Definition~\ref{Def1} below.

\begin{remark}\label{rem2}
In equation (\ref{vij}), the matrices $H_{ij}\in \mathbf{R}^{r_i\times n}$
and $G_{ij}\in \mathbf{R}^{r_i\times s_{ij}}$ do not depend on
  $\eta_i(t)$. This is
to reflect a situation where  
node $j$ \emph{always} broadcasts its information to node $i$, but node $i$
randomly fails to receive this information, or chooses not to accept it,
e.g. due to random congestion. 
It is possible to consider a
more general situation where the matrices $H_{ij}$ and $G_{ij}$ also depend on
$\eta_i(t)$. Technically, this more general case is no different from the one
pursued here.
\end{remark}

Associated with the system (\ref{eq:plant.1}) and the set of filters
(\ref{UP7.C.d.loc}) is
the disagreement function~(cf.~\cite{OM-2004})  
\begin{eqnarray}\label{disagr}
\Psi^k(\hat x)=\frac{1}{N} \sum_{i=1}^{N} \sum_{j\in
  \mathbf{V}^{\Phi_i(k)}}\|\hat x_j-\hat x_i\|^2, \quad 
k\in\mathcal{I}, 
\end{eqnarray}
$\hat x\triangleq [\hat x_1'~\ldots~\hat x_{N}']'$. 
It represents the average (over the set of all
nodes) of the total disagreement between the estimate at each node, and the
estimates computed at the neighbouring nodes, when the network is in state $k$. 
Following~\cite{U6}, we adopt $\Psi^k(\hat x)$ to define
the transient consensus performance metric in the distributed estimation problem
defined below.   

Let $\mathsf{P}^{x_0,m_0}$, $\mathsf{E}^{x_0,m_0}$ denote
the conditional probability and conditional expectation, given $x(0)-\hat
x_i(0)=x_0$ $\forall i$, $\eta(0)=m_0$.  Also, given a matrix $P=P'>0$, let
\begin{eqnarray*}
&&\mu_P(x_0,\xi,[\xi_i,w_{ij}]_{i,j=1,\ldots,N})\triangleq 
\|x_0\|^2_P + \|\xi\|_2^2 \nonumber\\
&&\qquad 
+\frac{1}{N} \sum_{i=1}^N\bigg(\|\xi_i\|_2^2
+\sum_{j=1}^M\mathsf{E}^{x_0,m_0}\|\mathbf{a}_{ij}^{\eta(\cdot)} w_{ij}\|^2_2\bigg). 
\nonumber
\end{eqnarray*}

\begin{definition}\label{Def1}
The distributed estimation problem under consideration is
to determine switching observer gains
${\tilde L}_i^k$ and interconnection coupling gains
${\tilde K}_{ij}^k,~k\in\mathcal{I}_i$, for the filters 
(\ref{UP7.C.d.loc}) which ensure that the following conditions are
satisfied:
  \begin{enumerate}[(i)]
  \item
In the absence of the uncertainty, all node estimators converge
exponentially in the mean-square sense and converge asymptotically with probability~1:
\begin{eqnarray*}
&&
\mathsf{E}^{x_0,m_0} \|\hat x_i(t)-x(t)\|^2
\le c e^{-\epsilon t}, \quad (\exists c,\epsilon>0), \\
&& 
\mathsf{P}^{x_0,m_0} (\lim_{t\to\infty}\|\hat x_i(t)-x(t)\|^2= 0)=1.
\end{eqnarray*}

\item Given a constant $\gamma>0$, 
the following mean-square $H_\infty$ consensus performance is guaranteed
\begin{eqnarray}\label{objective.i.1}
&&\sup_{x_0, (\xi,\xi_i,w_{ij})\neq 0}\,
\frac{\mathsf{E}^{x_0,m_0}\int_0^\infty\Psi^{\eta(t)}(\hat x(t))dt}
{\mu_P(x_0,\xi,[\xi_i,w_{ij}]_{i,j=1,\ldots,N})}
\le \gamma^2.
\end{eqnarray}

\item
 All estimators converge in the mean-square and with probability 1:
\begin{eqnarray}
&&\mathsf{E}^{x_0,m_0}\int_0^\infty \|x(t)-\hat x_i(t)\|^2dt <\infty,  
\label{convergence} \\
&&\mathsf{P}^{x_0,m_0} (\lim_{t\to\infty}\|x(t)-\hat x_i(t)\|^2= 0)=1.
 \label{convergence.P1}
\end{eqnarray}
  \end{enumerate}
\end{definition}
Properties (\ref{convergence}) and (\ref{convergence.P1}) refer to
different types of asymptotic behaviour of the estimation errors. Condition
(\ref{convergence}) states that $\hat x_i(t)$ converges to $x(t)$
in the mean-square $L_2$ sense. From the Chebyshev inequality, this also implies that
$\lim_{R\to \infty}\mathsf{P}^{x_0,m_0}
\left(\int_0^\infty\|x(t)-\hat x_i(t)\|^2dt> R\right)=0$,
that is, almost all estimator trajectories converge in $L_2$ sense.
Property (\ref{convergence.P1}) states that $\|x(t)-\hat x_i(t)\|^2$ converges to zero
asymptotically for almost
all realizations of the global state process $\eta(t)$. This is a stronger
property; in general, it does not follow from the a.s.~$L_2$
convergence. For that reason, both convergence properties are
considered in Definition~\ref{Def1}.    

\section{An auxiliary global distributed estimation
  problem}\label{sec:design-global} 

\subsection{Non-fragile distributed estimation}  

In this section, we temporarily lift the locality information structure
constraint and assume the
global communication and sensing state process $\eta(t)$ to be available at
every node. 

For every $k\in\mathcal{I}$ define $C_i^k=\tilde C_i^{\Phi_i(k)}$
$D_i^k=\tilde D_i^{\Phi_i(k)}$, $ 
\bar D_i^k=\tilde {\bar D}_i^{\Phi_i(k)}$. Note that $E_i^k\triangleq D_i^k(
 D_i^k)'+\bar D_i^k(\bar D_i^k)'=\tilde E_i^{\Phi_i(k)}>0$.
Then, the measurements taken at node $i$ can be rewritten in terms of
the global state process $\eta(t)$:  
\begin{equation}\label{U6.yi.2}
y_i=C_i^{\eta(t)}x+D_i^{\eta(t)}\xi+\bar  D_i^{\eta(t)}\xi_i.
\end{equation}

The auxiliary problem in this section is concerned with estimation of the
state of the uncertain plant (\ref{eq:plant.1}), (\ref{U6.yi.2}) using a
network of estimators subject to uncertainty, as follows 
\begin{eqnarray}
    \dot{\hat x}_i&=&A\hat x_i +
    L_i^{\eta(t)}(y_i(t)-C_i(\eta(t))\hat x_i)\nonumber \\
    &+&\sum_{j\in
      \mathbf{V}^{\Phi_i(\eta(t))}} K_{ij}^{\eta(t)}(v_{ij}-H_{ij}\hat
    x_i) \nonumber \\
&+&\sum_{j\in
      \mathbf{V}^{\Phi_i(\eta(t))}}(\omega_{ij}^{(1)}+\omega_{ij}^{(2)})+\omega_i,  
  \label{UP7.C.d.2} \quad \hat x_i(0)=0. \quad
\end{eqnarray}
Here, $L_i^{(\cdot)}$, $K_{ij}^{(\cdot)}$  are matrix-valued functions of the 
state of the global Markov chain $\eta$ to be found, and $\omega_{ij}^{(1)}$,
$\omega_{ij}^{(2)}$, and $\omega_i$ are estimator perturbations. It is
assumed that these perturbations are random processes adapted to the filtration
$\{\mathcal{F}_t,t\ge 0\}$ and such that the multivariate process 
$(\hat x_1,\ldots, \hat x_N,\eta)$ is Markov with respect to that
filtration. Also in this section, it will be assumed that these
uncertainties satisfy the following norm-bound conditions:
\begin{eqnarray}
&&\|\omega_i(t)\|^2 \le \alpha_i^2
\left\|C_i^{\eta(t)}e_i(t)+D_i^{\eta(t)}\xi(t)+\bar
  D_i^{\eta(t)}\xi_i(t)\right\|^2, \nonumber \\ 
&&\|\omega_{ij}^{(1)}(t)\|^2\le \beta_{ij}^2
\left\|H_{ij}e_i(t)+G_{ij}w_{ij}\right\|^2, \nonumber \\
&&\|\omega_{ij}^{(2)}(t)\|^2\le \beta_{ij}^2
\left\|H_{ij}e_j(t)\right\|^2 \quad \mbox{a.s. } \forall t\ge 0,  \label{wv.constr} 
\end{eqnarray}
where $\alpha_i$, $\beta_{ij}$ are given constants, and  
$e_i=x-\hat x_i$ is the estimation error of the auxiliary estimator 
at node $i$, which evolves 
according to the
equations 
 \begin{eqnarray}
    \dot{e}_i&=&(A - L_i^{\eta(t)}C_i^{\eta(t)})e_i\nonumber \\
    &+& \sum_{j\in 
      \mathbf{V}_i^{\Phi_i(\eta(t))}} K_{ij}^{\eta(t)}(H_{ij}(e_j-e_i)-G_{ij}w_{ij}) \nonumber \\
    &+&(B_2-L_i^{\eta(t)}D_i^{\eta(t)})\xi- L_i^{\eta(t)}{\bar
      D_i^{\eta(t)}}\xi_i \nonumber \\
&-& \left(\sum_{j\in
      \mathbf{V}^{\Phi_i(\eta(t))}}(w_{ij}^{(1)}+w_{ij}^{(2)})+w_i\right),
  \quad e_i(0)=x_0.   \label{e.w} 
\end{eqnarray}

It will be shown in Section~\ref{main}
that when the locality information structure constraints are imposed,
this will result in an uncertainty due to the mismatch between filter error
dynamics in the network subject to these constraints, and 
the errors which would arise in the same network if its communication state was
known globally. It will be shown
in the proof of 
Theorem~\ref{T.main} that this uncertainty satisfies conditions
(\ref{wv.constr}). The resilience of the constraint-free auxiliary network 
(\ref{UP7.C.d.2}) to this uncertainty will be used in the next section to
show that the
network~(\ref{UP7.C.d.loc}), constructed from the auxiliary solution,
maintains the same convergence and robust $H_\infty$ consensus performance
properties when the information structure constraints are enforced.    

\begin{definition}\label{Def1.aux}
The auxiliary distributed consensus estimation problem is to
determine sets of gains
$L_i^k$, 
$K_{ij}^k$, $k\in\mathcal{I}$, for the filters 
(\ref{UP7.C.d.2}) to ensure the following:
\begin{enumerate}[(i)]
  \item
When $(\xi,\xi_i,w_{ij})\equiv 0$, the interconnected system consisting of
subsystems (\ref{e.w}) must be exponentially stable in the mean-square
sense and asymptotically 
stable with probability 1 for all estimator perturbations $\omega_{ij}^{(1)}$,
$\omega_{ij}^{(2)}$, and $\omega_i$ for which the correspondingly modified
constraints (\ref{wv.constr}) hold. 

\item
In the presence of exogenous disturbances $\xi,\xi_i,w_{ij}$, the
mean-square consensus 
performance condition in (\ref{objective.i.1}) is satisfied for all
admissible estimator perturbations  $\omega_{ij}^{(1)}$,
$\omega_{ij}^{(2)}$, and $\omega_i$ subject to (\ref{wv.constr}).

\item
 All estimators converge in the mean-square and with probability 1; i.e.,
 conditions (\ref{convergence}), (\ref{convergence.P1}) hold. 
\end{enumerate} 
\end{definition}

A solution to this auxiliary problem is
given in Theorem~\ref{T.aux} below.  The conditions of
the theorem involve the following linear matrix inequalities in the variables
$\tau_i^k>0$, $\theta_{ij}^k>0$, $\vartheta_{ij}^k>0$, $X_i^k=(X_i^k)'>0$,
$i=1,\ldots,N$, $k\in\mathcal{I}$, $j\in \mathbf{V}_i^{\Phi_i(k)}$:   
\begin{eqnarray}
  \label{T4.1}
&&\gamma^2I-\tau_i^k\alpha_i^2E_i^k>0, \quad \gamma^2I-\theta_{ij}^k\beta_{ij}^2
F_{ij}>0, \\
&&\left[
\begin{array}{ccccc}
Q_i^k & \star & \star & \star & \star \\
N_i^k & -\gamma^2I & \star &
\star & \star \\
S_i^k & 0 & -\gamma^2 I & \star & \star \\
\mathbf{1}_{1+2M_i} \otimes X_i^k & 0  & 0 & -\mathbf{T}_i & \star \\
\Xi_i' & 0  & 0 & 0 & -Z_i\\
\end{array}
\right]<0, \label{T4.LMI.1} \qquad 
\end{eqnarray}
where 
\begin{eqnarray*}
&&
N_i^k\triangleq
  \left(I-(D_i^k)'(E_i^k)^{-1}D_i^k\right)B_2'X_i^k, \\
&&
S_i^k\triangleq -(\bar D_i^k)'(E_i^k)^{-1}D_i^k B_2'X_i^k,\\
&& 
\mathbf{T}_i^k\triangleq \mathrm{diag}\left[\tau_i^k,~\theta_{i,j_1}^k,~\ldots,~\theta_{i,j_{p_i^k}}^k,
~\vartheta_{i,j_1}^k,~\ldots,~\vartheta_{i,j_{p_i^k}}^k\right], \\
&&
Q_i^k\triangleq X_i^k(A+\delta_iI-B_2(D_i^k)'(E_i^k)^{-1}C_i^k)\\
&&\phantom{Q_i^k}+(A+\delta_iI-
B_2(D_i^k)'(E_i^k)^{-1}C_i^k)'X_i^k  +(p_i^k+q_i^k)I \\
&& \phantom{Q_i^k}
+\sum_{j: i\in
  \mathbf{V}_j^{\Phi_j(k)}}\vartheta_{ji}^k\beta_{ji}^2H_{ji}'H_{ji}+\sum_{l=1}^M\lambda_{kl}X_i^l  \\
&& \phantom{Q_i^k}
-\gamma^2(C_i^k)'(E_i^k)^{-1}C_i^k- \gamma^2\sum_{j\in
  \mathbf{V}_i^{\Phi_i(k)}}H_{ij}'F_{ij}^{-1}H_{ij},\\
&&
\Xi_i\!=\!\left[\begin{array}{ccc}\gamma^2H_{ij_1}'F_{ij_1}^{-1}H_{ij_1}-I & \ldots & \gamma^2H_{ij_{p_i^k}}'F_{ij_{p_i^k}}^{-1}H_{ij_{p_i^k}}-I
  \end{array}
  \right]\!, \\
&&
Z_i=\mathrm{diag}\left[\frac{2\delta_{j_1}}{q_{j_1}^k+1} X_{j_1}^k,~\ldots,
  ~ \frac{2\delta_{j_{p_i^k}}}{q_{j_{p_i^k}}^k+1} X_{j_{p_i^k}}^k \right].
\end{eqnarray*}

\begin{theorem}\label{T.aux}
Suppose the network $(\eta,\mathcal{G},\Phi)$ and the constants $\gamma>0$,
$\alpha_i$, $\beta_{ij}$ and $\delta_i>0$ are such that 
the coupled LMIs (\ref{T4.1}) and (\ref{T4.LMI.1})
in the variables $\tau_i^k>0$, $\theta_{ij}^k>0$, $\vartheta_{ij}^k>0$,
$X_i^k=(X_i^k)'>0$, $j\in \mathbf{V}_i^{\Phi_i(k)}$,  
$i=1,\ldots,N$, $k\in\mathcal{I}$, are feasible. 
Then the network of observers (\ref{UP7.C.d.2}) with
\begin{eqnarray}
 K_{ij}^k&=&\gamma^2(X_i^k)^{-1}H_{ij}'F_{ij}^{-1}, \label{Kim} \nonumber \\
 L_i^k&=&\left[\gamma^2(X_i^k)^{-1}(C_i^k)'+
B_2 (D_i^k)'\right](E_i^k)^{-1} \label{Lim}
\end{eqnarray}
solves the auxiliary estimation problem in Definition~\ref{Def1.aux}. The
matrix $P$ in condition (\ref{objective.i.1}) 
corresponding to this solution is $P=\frac{1}{\gamma^2N}\sum_{i=1}^NX_i^{m_0}$,
where $m_0=\eta(0)$.  
\end{theorem}

The proof of Theorem~\ref{T.aux} is given in the Appendix. 

\subsection{Special case: Broadcast of the global state}   

When the global Markov state of the network is available at
every node, the solution to the distributed
$H_\infty$ consensus estimation problem can be obtained from
Theorem~\ref{T.aux} by letting  
$\omega_i=\omega_{ij}^{(1)}=\omega_{ij}^{(2)}=0$ and taking
$\alpha_i=\beta_{ij}=0$.

\begin{corollary}\label{Corollary}
Suppose the network $(\eta,\mathcal{G},\Phi)$ and the constants $\gamma>0$
and $\delta_i>0$ are such that the following coupled LMIs in the variables 
$X_i^k=(X_i^k)'>0$, $j\in \mathbf{V}_i^{\Phi_i(k)}$,  
$i=1,\ldots,N$, $k\in\mathcal{I}$, are feasible:
\begin{eqnarray}
\lefteqn{\left[
\begin{array}{cccc}
\bar Q_i^k & \star & \star & \star \\
N_i^k & -\gamma^2I & \star & \star \\
S_i^k & 0 & -\gamma^2 I & \star \\
\Xi_i' & 0  & 0 & -Z_i\\
\end{array}
\right]<0,} && \label{T4.LMI.Cor} 
\\
\lefteqn{\bar Q_i^k\triangleq
X_i^k(A+\delta_iI-B_2(D_i^k)'(E_i^k)^{-1}C_i^k)} && \nonumber \\
&&+(A+\delta_iI-
B_2(D_i^k)'(E_i^k)^{-1}C_i^k)'X_i^k  +(p_i^k+q_i^k)I \nonumber\\
&&
+\sum_{l=1}^M\lambda_{kl}X_i^l
-\gamma^2(C_i^k)'(E_i^k)^{-1}C_i^k- \gamma^2\hspace{-2ex}\sum_{j\in
  \mathbf{V}_i^{\Phi_i(k)}}H_{ij}'F_{ij}^{-1}H_{ij}. \nonumber
\end{eqnarray}
Then the network of observers (\ref{UP7.C.d.2}) with
$\omega_i=\omega_{ij}^{(1)}=\omega_{ij}^{(2)}=0$ and $K_{ij}^k$, $L_i^k$
defined in (\ref{Lim})
solves the estimation problem in Definition~\ref{Def1.aux}. The
matrix $P$ in condition (\ref{objective.i.1})
corresponding to this solution is $P=\frac{1}{N\gamma^2}\sum_{i=1}^NX_i^{m_0}$,
where $m_0=\eta(0)$.  
\end{corollary}

\section{The main result}\label{main}

In this section, the solution to the auxiliary distributed estimation
problem developed in Section~\ref{sec:design-global}  will be used to
obtain a distributed estimator whose nodes utilize only locally available 
information. This will be achieved by  
taking the asymptotic conditional expectation of the auxiliary gains, given 
a local state. 
Our method is based on the following
technical result of \cite{XiUP1}.

\begin{lemma}
  \label{prop:1}
  Suppose the Markov process $\eta(t)$ is irreducible and has a unique
  invariant distribution $\bar\lambda$. Given a matrix-valued function
  $K^{(\cdot)}:\mathcal{I}\to \{K^1\ldots, K^M\}\subset 
  \mathbf{R}^{n\times s}$, for every node $i$ and for all
  $k_i\in\mathcal{I}_{i}$  we have:  
  \begin{eqnarray}
    \lim_{t\to\infty} \mathsf{E}\left(K^{\eta(t)} \mid \eta_{i}(t) = k_i\right)
= \frac{\sum_{l:\Phi_i(l)=k_i}
        \bar\lambda_lK^l}
    {\sum_{l:\Phi_i(l)=k_i}\bar\lambda_l}.
  \end{eqnarray}
\end{lemma}

Now let $K_{ij}^k$, $L_i^k$, $k\in\mathcal{I}$, be the coefficients of the
auxiliary distributed estimator obtained in
Theorem~\ref{T.aux}. Using Lemma~\ref{prop:1}, for each $i=1,\ldots,
N$ and $k_i\in\mathcal{I}_i$ we define 
  \begin{eqnarray}
    \tilde K_{ij}^{k_i}    = \frac{\sum_{l:\Phi_i(l)=k_i}
        \bar\lambda_lK_{ij}^l}
    {\sum_{l:\Phi_i(l)=k_i}\bar\lambda_l},\label{Kim.tilde}
\quad
    \tilde L_i^{k_i}= \frac{\sum_{l:\Phi_i(l)=k_i} \bar\lambda_l L_i^l}
    {\sum_{l:\Phi_i(l)=k_i}\bar\lambda_l}. \qquad
 \label{Lim.tilde} 
  \end{eqnarray}
From Lemma~\ref{prop:1}, the processes $\tilde K_{ij}^{\eta_i(t)},
\tilde L_i^{\eta_i(t)}$  are then
the asymptotic minimum variance approximations of the corresponding processes
$K_{ij}^{\eta(t)}, L_i^{\eta(t)}$. 
However, unlike $K_{ij}^{\eta(t)}, L_i^{\eta(t)}$, the
evolution of $\tilde
K_{ij}^{\eta_i(t)}, \tilde L_i^{\eta_i(t)}$ is governed by the local
communication and sensing state process $\eta_i$.  

To formulate the main result of this paper, 
consider the collection of the LMIs in the variables
$\tau_i^k$, $\theta_{ij}^k$, $\vartheta_{ij}^k$, $X_i^k$ and $Y_i^k$,
consisting of the LMIs (\ref{T4.1}), (\ref{T4.LMI.1}), and the following
additional LMIs, 
\begin{eqnarray}
  \label{DeltaL}
&&  \left[
    \begin{array}{cc}
      \alpha_i^2 I & \Delta_i^{L,k} \\ (\Delta_i^{L,k})' & I
    \end{array}
\right]>0, \quad 
  \label{DeltaK}
  \left[
    \begin{array}{cc}
      \beta_{ij}^2 I & \Delta_{ij}^{K,k} \\ (\Delta_{ij}^{K,k})' & I
    \end{array}
\right]>0,
\end{eqnarray}
where $\alpha_i$, $\beta_{ij}$ are the same constants as those employed in
the LMIs (\ref{T4.1}), (\ref{T4.LMI.1}), and
\begin{eqnarray*}
\Delta_i^{L,k}&\triangleq &\frac{\sum\limits_{l:l\neq k,\atop \Phi_i(l)=\Phi_i(k)}\!\!\!\!\!
      \gamma^2\bar\lambda_l
      \left[Y_i^k(C_i^k)'(E_i^k)^{-1}-Y_i^l(C_i^l)'(E_i^l)^{-1}\right]}{\sum_{l:\Phi_i(l)=\Phi_i(k)}
      \bar\lambda_l } , \\[1ex]
\Delta_{ij}^{K,k}&\triangleq &\frac{\sum\limits_{l:l\neq k,
    \atop\Phi_i(l)=\Phi_i(k)}\!\!\!\!\!
\gamma^2\bar\lambda_l
      \left[Y_i^k-Y_i^l\right]H_{ij}'F_{ij}^{-1}}{\sum_{l:\Phi_i(l)=\Phi_i(k)}
      \bar\lambda_l } 
      .
\end{eqnarray*}
Also, consider the rank constraints
\begin{eqnarray}
&&\mathrm{rank}\left[\begin{array}{cc} Y_i^k & I \\ I & X_i^k
  \end{array}
\right] \le n, 
  \label{rank}
\end{eqnarray}

\begin{theorem}\label{T.main}
Given a Markovian switching network $(\eta,\mathcal{G},\Phi)$ and a
collection of 
constants $\gamma$, $\alpha_i$, $\beta_{ij}$ and $\delta_i>0$, $i=1,\ldots,
N$, associated with each node, suppose
there exist matrices $X_i^k=(X_i^k)'>0$,  
$Y_i^k=(Y_i^k)'>0$, and positive scalars  
  $\tau_i^k$, $\theta_{ij}^k$, $\vartheta_{ij}^k$, $i=1,\ldots,N$,
  $k\in\mathcal{I}$, $j\in\mathbf{V}_i^{\Phi_i(k)}$ which satisfy the
  matrix inequalities (\ref{T4.1}), (\ref{T4.LMI.1}), (\ref{DeltaL}),
  and the rank constraint (\ref{rank}).
  Using the solution matrices $Y_{i}^k$, define the auxiliary gains 
\begin{eqnarray}
 K_{ij}^k&=&\gamma^2Y_i^kH_{ij}'F_{ij}^{-1},  \label{Kim.Y} \nonumber
 \\
 L_i^k&=&\left[\gamma^2Y_i^k(C_i^k)'+
B_2 (D_i^k)'\right](E_i^k)^{-1}. \label{Lim.Y}
\end{eqnarray}
Next, using~(\ref{Kim.tilde}) and 
(\ref{Lim.Y}), 
  construct the estimator network (\ref{UP7.C.d.loc}). The resulting distributed
  estimatior network solves the distributed robust estimation problem in
  Definition~\ref{Def1}.  
\end{theorem}

\emph{Proof }
The result follows from Theorem~\ref{T.aux} in a manner similar to the proof of
Theorem~4 in~\cite{XiUP1}.

First we observe that the observer gains $K_{ij}^k$, $L_i^k$ constructed in
this theorem, also satisfy the conditions of Theorem~\ref{T.aux}, since
$(X_i^k)^{-1}=Y_i^k$ in view of (\ref{rank}). This allows us to claim that 
the network of auxiliary estimators
(\ref{UP7.C.d.2}), (\ref{Lim.Y}) solves the auxiliary robust estimation problem in
Definition~\ref{Def1.aux}.

Next, consider the observer gains
defined using~(\ref{Kim.tilde}) and (\ref{Lim.Y}). 
Note that for all $i=1,\ldots,N$, $k\in\mathcal{I} $,
and $j\in\mathbf{V}_i^{\Phi_i(k)}$,  
\begin{eqnarray*}
   \label{eq:32.K}
  K_{ij}^k - \tilde K_{ij}^{\Phi_i(k)} &=&
  \frac{\sum_{l:l\neq k, \Phi_i(l)=\Phi_i(k)}  \bar\lambda_l 
\left[K_{ij}^k-K_{ij}^l\right]}{\sum_{l:\Phi_i(l)=\Phi_i(k)}
\bar\lambda_l}, \\
  \label{eq:32.L}
  L_i^k - \tilde L_i^{\Phi_i(k)} &=&
  \frac{\sum_{l:l\neq k, \Phi_i(l)=\Phi_i(k)}
      \bar\lambda_l \left[L_i^k-L_i^l\right]} {\sum_{l:\Phi_i(l)=\Phi_i(k)}
      \bar\lambda_l },
\end{eqnarray*}
  Then it follows from (\ref{DeltaK}) that 
\begin{eqnarray}
\|\tilde L_i^{\Phi_i(k)}-L_i^k\|^2\le \alpha_i^2, \quad
\|\tilde K_{ij}^{\Phi_i(k)}-K_{ij}^k\|^2\le \beta_{ij}^2. 
\label{wv.constr.nb}
\end{eqnarray}
Therefore the particular perturbations in the
estimators (\ref{UP7.C.d.2}),
 \begin{eqnarray}
   \omega_i   &=&(\tilde
    L_i^{\eta_i(t)}-L_i^{\eta(t)})(C_i^{\eta(t)}e_i(t)+D_i^{\eta(t)}\xi+\bar
    D_i^{\eta(t)}\xi_i), \nonumber \\
   \omega_{ij}^{(1)} &=&
(\tilde K_{ij}^{\eta_i(t)}-K_{ij}^{\eta(t)})(H_{ij}e_i+G_{ij}w_{ij}), 
\nonumber \\
   \omega_{ij}^{(2)}&=&-
(\tilde K_{ij}^{\eta_i(t)}-K_{ij}^{\eta(t)})H_{ij}e_j,
\label{wv}
 \end{eqnarray}
satisfy  (\ref{wv.constr}). 
This means that the estimator (\ref{UP7.C.d.loc}) in which the above
particular set of gains $\tilde K_{ij}^{\eta_i(t)}, \tilde L_i^{\eta_i(t)}$
is employed, represents one instance of the auxiliary  
estimator (\ref{UP7.C.d.2}), corresponding to the particular perturbation
(\ref{wv}), which is an admissible perturbation, due to (\ref{wv.constr}). 
Therefore, since the matrices $K_i^k$, $L_i^k$, 
$m\in\mathcal{I}$ solve the auxiliary $H_\infty$ consensus estimation
problem in Definition~\ref{Def1.aux}, then the distributed estimator
(\ref{UP7.C.d.loc}) with the local gains selected above, solves the robust
consensus estimation problem in Definition~\ref{Def1}. \hfill$\Box$

\begin{remark}
  Due to the rank constraints \eqref{rank}, the solution set to the matrix
  inequalities in Theorem~\ref{T.main} is non-convex. In general, it is
  difficult to solve such problems. Fortunately, several numerical
  algorithms have been proposed for this purpose
\cite{EOA-1997,OHM-2006}.
\end{remark}

\section{Requirements on the communication graph and
  interconnections}\label{requirements}

In this section, we briefly discuss necessary requirements on the network topology. Recall that condition (i) of Definition~\ref{Def1} requires
that in the absence of perturbations, estimation error dynamics must be
asymptotically stabilizable via output injection in the mean-square.   
This problem belongs to the class of stochastic mean-square detectability
problems for linear jump parameter systems~\cite{CdV-2002}. 
Unfortunately, even without the locality information structure constraint,
there is no easy  direct
algebraic test to verify this property. Some conclusions
can however be drawn to provide an insight
into the connection between the graph Laplacian and the existence of
stabilizing output injection gains. 

To highlight such a connection, in this section we
will make the simplifying assumption $H_{ij}=H$, $G_{ij}=G_i$, $r_{ij}=\bar
r_i$ for all $j\in
\mathbf{V}_i^{\Phi_i(k)}$. From (\ref{Kim.Y}), it follows that in this case
$\tilde K_{ij}^{k_i}$ does not depend on $j$. Hence
we will also assume 
$\tilde K_{ij}^{k_i}=\tilde K_i^{k_i}$. Define
$\bar A=I_N\otimes A$, $A_k\triangleq A+\frac{1}{2}\lambda_{kk}I$, 
$\bar A_k\triangleq \bar A+\frac{1}{2}\lambda_{kk}I_{nN}$, $\bar
H^k=\mathcal{L}^k\otimes H$. 

Let $\mathcal{C}_i^k$, $\bar{\mathcal{O}}^k$, $\mathcal{O}_H$ denote the 
undetectable subspace of $(C_i^k, A_k)$ and the unobservable subspaces of
$(\bar H^k, \bar A_k)$ and $(H, A)$, respectively. The following theorem shows that for the problem in
Definition~\ref{Def1} to have a solution, every combination of undetectable
states of the pairs $(C_i^k, A_k)$ must necessarily form an observable
vector of $(\bar H^k, \bar A_k)$. 
The proofs of this and subsequent results are removed for the sake of brevity.

\begin{theorem}\label{U7.prop1}
Suppose there exist output injection gains 
${\tilde L}_i^{k_i}$, ${\tilde K}_{ij}^{k_i}={\tilde K}_i^{k_i}$, $j\in
\mathbf{V}_i^{k_i}$, 
$k_i\in\mathcal{I}_i$, $i=1,\ldots,N$, 
such that the first condition in Definition~\ref{Def1}(i) holds. Then, 
\begin{equation}
  \label{undetect.C}
\bar{\mathcal{O}}^k\cap \prod_{i=1}^N\mathcal{C}_i^k =\{0\} \quad \forall k\in 
\mathcal{I}.
\end{equation}
\end{theorem}

We now present a necessary and sufficient condition for (\ref{undetect.C})
to hold. The sufficient condition is explicitly expressed in terms of the
network Laplacian matrices $\mathcal{L}^k$.   

\begin{theorem}\label{U7.prop.2}
\begin{enumerate}[(a)]
\item
If (\ref{undetect.C}) holds, then 
for every $k\in\mathcal{I}$:
\begin{enumerate}[(i)]
\item
$\bigcap_{i=1}^N \mathcal{C}_i^k=\{0\}$, and
\item 
$\mathcal{O}_H\cap \mathcal{C}_i^k=\{0\}$ for all $i=1,\ldots,N$;
\end{enumerate}
\item
Conversely, suppose the geometric multiplicity of
the zero eigenvalue of $\mathcal{L}_k$ is equal to
1. If the above properties (i) and (ii) hold for every
$k$, then (\ref{undetect.C}) is satisfied. 
\end{enumerate}
\end{theorem}  

One can further specialize the sufficient conditions in Theorem 4, e.g.,
for the cases of a balanced graph or a graph containing a spanning tree. 
Also, note that the information structure constraint is not used
in the proofs. Therefore, the results in this section apply to more general
distributed estimation problems, such as the auxiliary problem
considered in Section~\ref{sec:design-global}.

\section{Example}\label{Example}
%
Consider a plant of the form (\ref{eq:plant.1}), with 
$
A=\left[\begin{smallmatrix}
   -3.2 &  10    &  0\\
    1   &  -1    &  1 \\
    0   & -14.87 &  0
  \end{smallmatrix}
\right]$, \quad
$B_2=\left[\begin{smallmatrix}
    -0.1246 \\
   -0.4461 \\
    0.3350 
  \end{smallmatrix}
\right]$.
The nominal part of the plant describes one of the regimes of the so-called
Chua electronic circuit. The Chua circuit is an example of a system which
switches between three regimes of operation in a chaotic
fashion. For the sake of simplicity, here we consider only one regime.

\begin{figure}
  \psfrag{1}{1}
  \psfrag{2}{2}
  \psfrag{3}{3}
  \psfrag{4}{4}
  \psfrag{5}{5}
  \psfrag{m=1}{$k=1$}
  \psfrag{m=2}{$k=2$}
\centering
  \includegraphics[width=0.9\columnwidth]{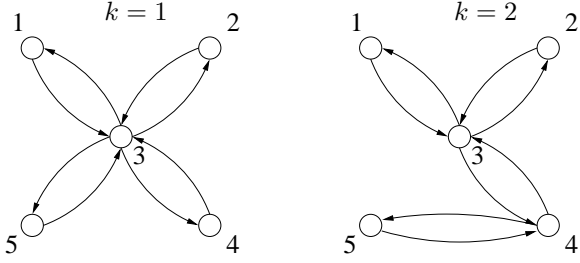}
  \caption{Switching graph topology for the example.}
  \label{fig.ex}
\end{figure}

The plant is observed by a 5-node switching observer network which
operates intermittently in two regimes. Its graph topologies are shown in
Figure~\ref{fig.ex}. The evolution of the network is modelled as
a two-state Markov chain with the transition probability rate matrix
$\Lambda=\left[\begin{smallmatrix}
-0.1 & 0.1 \\ 0.1 & -0.1
  \end{smallmatrix}
\right]$.
Note that the graph corresponding to state $k=2$ was used in
\cite{U6,YGH-2009} to demonstrate synchronization of Chua systems. Indeed,
the filters share the same 
matrix $A$ as the plant, and can be interpreted as `slave' Chua systems operating in the same regime as the master. Accordingly, the convergence of the filters
in our example can be interpreted as the observer-based
synchronization between the slaves and the master; see \cite{U6} for further
details. However different from \cite{U6}, in this example the graph topology
is time-varying, as explained below.

From Figure~\ref{fig.ex}, nodes 3, 4, and
5 have varying neighbourhoods. Also, in this example we suppose that node 2
changes its sensor parameters when the network switches between two
configurations. As a result, in this example, each local state process,
except for that of node 1, has two states and always takes the same value as
the global state process. On the other hand, node 1 always maintains the same
local state, and its local process is constant. Therefore,    
we seek to obtain nonswitching observer gains for node 1 only. According to
this description, in this example,
$\mathcal{I}=\mathcal{I}_2=\mathcal{I}_3=\mathcal{I}_4=\mathcal{I}_5=\{1,2\}$,
$\mathcal{I}_1=\{1\}$, and the mapping $\Phi$ is as follows:
$\Phi(1)=(1,1,1,1,1)$, $\Phi(2)=(1,2,2,2,2)$. 
Numerical values of the matrices $C_i^k$, $k=1,2$, for this example
are given in Table~\ref{TableC}; they are assumed to take one of the two
values $C_{*1}$, $C_{*2}$, shown in the table. These values were chosen so
that the pairs $(C_1^k, A+\frac{1}{2}\lambda_{kk}I)$,  $(C_4^k,
A+\frac{1}{2}\lambda_{kk}I)$, $k=1,2$, corresponding to nodes 1 and 4, had
undetectable modes, while node 2 was allowed to switch between detectable and
undetectable coefficient pairs. Therefore, for estimation these nodes were to
rely on communication with their neighbours.  
Also,    
we let $D_i^k=0$, $\bar D_i^k=0.025$ for all nodes and all $k$, and
$H_{ij}=I_{3\times 3}$, $G_{ij}=0.5\times I_{3\times 3}$. 

Note that both instances of the network have 
spanning trees with roots at nodes 3 and 5. These
nodes have detectable matrix pairs $(C_3^k, A+\frac{1}{2}\lambda_{kk}I)$,
$(C_5^k, A+\frac{1}{2}\lambda_{kk}I)$, $k=1,2$, respectively. Also,
$(H,A)$ is observable. It follows from these
properties that the conditions in part (b)
of Theorem~\ref{U7.prop.2} are satisfied. Hence, the
necessary condition for 
 global detectability, stated in Theorem~\ref{U7.prop1} holds.
\begin{table}
  \caption{Coefficients $C_i^k$ for the example, $C_{*1}=10^{-3}\times [3.1923~
-4.6597~    1]$, $C_{*2}=[-0.8986~    0.1312~ -1.9703]$.} \label{TableC}
  \centering
  \begin{tabular}{|c|c|c|c|c|c|}
\hline 
       & $i=1$ & $i=2$ & $i=3$ & $i=4$ & $i=5$  \\ \hline
$k=1$  & $C_{*1}$ & $C_{*1}$ & $C_{*2}$ & $C_{*1}$ & $C_{*2}$ \\
$k=2$  & $C_{*1}$ & $C_{*2}$ & $C_{*2}$ & $C_{*1}$ & $C_{*2}$ \\
\hline
  \end{tabular}
\end{table}

The design of the observer network was carried out using Matlab and the
LMI solver LMIrank based on~\cite{OHM-2006}. To obtain a
set of non-switching gains for node 1, the
norm-bounded uncertainty constraints of the form 
(\ref{wv.constr.nb}) were defined 
for the communication link $(3,1)$ at node 1, where
we set $\alpha_{13}^2=10^2$, $\beta_{13}^2=4\times 10^2$. These constants
as well as $\delta_i=0.365$ were chosen by trial and error, to ensure that the
corresponding rank constrained LMIs in Theorem~\ref{T.main} were
feasible. The feasibility was achieved with
$\gamma^2=0.7407$. This allowed us to compute the nonswitching gains
$\tilde K_{13}$ and $\tilde L_1$  for node 1 using (\ref{Lim.Y}).   

\begin{figure}
\centering
\subfigure[\label{eta.graph}]{\includegraphics[width=0.9\columnwidth,height=3.7cm]{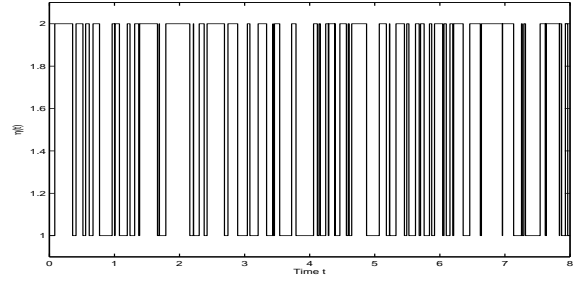}}
\subfigure[\label{errors.graph}]{\includegraphics[width=0.9\columnwidth,height=3.7cm]{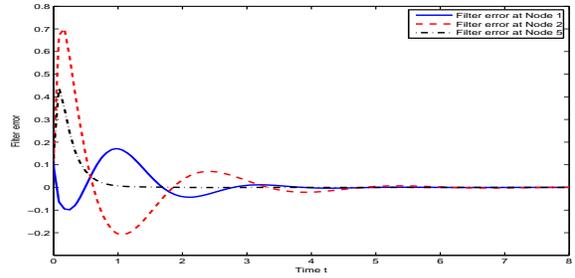}}
  \caption{One path of $\eta(t)$ (Fig.~(a)) and estimations errors for the
    first coordinate at nodes 1, 2 and 5 (Fig.~(b)).} 
\end{figure}

To validate the design, the system and the designed filters were
simulated numerically, with a random initial condition $x_0$. 
All uncertain perturbations were chosen to be of
the form $\sin(a\pi t+\varphi)e^{-bt}$, with different coefficients
$a,\varphi$ and $b$ for each perturbation. Also we let
$w_{ij}(t)=w_{ji}(t)$, assuming an undirected nature of the 
channels in this example.  
The graphs of one realization 
of the global state process $\eta(t)$, and the corresponding estimation
errors at nodes 1 (the nonswitching filter), 2 (the filter with the switching
sensing regime) and 5 (the filter with the varying neighbourhood) are
shown in Figures~\ref{eta.graph} and~\ref{errors.graph},
respectively. The graph in Figure~\ref{errors.graph} confirms the
ability of the 
proposed node estimators to successfully mitigate the 
changes in the graph topology and sensing
regimes, as well as uncertain perturbations in the plant, measurements and
interconnections. 

\section{Conclusions}\label{Conclusions}
The paper has presented sufficient conditions for the synthesis of robust
distributed consensus estimators connected over a Markovian network.
The proposed estimator provides a guaranteed suboptimal $H_\infty$
disagreement of estimates, while using only locally available information
about the communication and sensing state of the network. Our conditions allow a
robust filter network to be constructed by solving an LMI feasibility
problem. The LMIs are partitioned in a way which opens a possibility for
solving them in a decentralized manner. When the 
network's global state is available at every node, this feasibility problem is
convex, and the corresponding LMIs are solvable, e.g., using the decentralized
gradient descent algorithm in~\cite{U6}. However, the elimination
of the network state broadcast has led to the
introduction of rank constraints additional to the LMI
conditions. Therefore, new numerical algorithms need to be developed to
exploit the proposed partition of the LMIs and rank constraints. This
problem is left for future research.   
Other possible directions for future research may be concerned with 
an integration of our approach with other
distributed $H_\infty$ filtering techniques, such as for example,
techniques involving randomly sampled measurements~\cite{SWL-2011}. 

\section*{Acknowledgement}
Discussions with C.~Langbort are gratefully acknowledged. 

\section{Appendix: Proof of Theorem~\ref{T.aux}.}

The following continuous-time counterpart of the
Robbins-Siegmund convergence theorem~\cite{RS-1971} will be used in the proof of
Theorem~\ref{T.aux}. Its proof is similar to~\cite{RS-1971}.  

\begin{lemma}
  \label{Supermartingale.Lemma}
Consider nonnegative random processes $v(t)$, $\phi(t)$ and
$\psi(t)$ adapted to a filtration $\{\bar{\mathcal{F}}_t,t\ge 0\}$, with the
following properties:

\begin{enumerate}[(a)]
\item
$v(t)$ is right-continuous on $[0,\infty)$;
\item
$\psi(t)$ is locally Lebesgue-integrable on $[0,\infty)$ with probability
1, i.e., almost all realizations of $\psi(t)$ have the property $\int_s^t
\psi(\theta)d\theta <\infty$ for all $t\ge s\ge 0$;     
\item
$\mathsf{E}\int_0^\infty\phi(s)ds<\infty$;
\item
The following inequality holds a.s. for all $t\ge s\ge 0$
\begin{eqnarray}
  \label{supermart}
\hspace{-5ex}\mathsf{E}\left[v(t)+\int_s^t\psi(\theta)d\theta
    \big|\bar{\mathcal{F}}_s\right]\le v(s) +
  \mathsf{E}\left[\int_s^t\phi(\theta)d\theta\big|\bar{\mathcal{F}}_s\right].
\end{eqnarray}
\end{enumerate}
Then the limit $\lim_{t\to\infty} v(t)$ exists and is finite with
probability 1. Also, $\int_0^\infty\psi(s)ds<\infty$ a.s.. 
\end{lemma}

\paragraph*{Proof of Theorem~\ref{T.aux}}

We will use the notation
$k_i=\Phi_i(k)$, $k_j=\Phi_j(k)$, where $k\in \mathcal{I}$, $k_i\in
\mathcal{I}_i$, $k_j\in \mathcal{I}_j$. Also, 
 $\hat D_i^k=[\begin{array}{cc} D_i^k & \bar D_i^k
  \end{array}]$, 
$\hat B_2=[\begin{array}{cc} B_2 & 0
  \end{array}]$, $\hat \xi_i=[\begin{array}{c} \xi'~\xi_i'
  \end{array}]'$. 

Let $\mathfrak{L}$ denote the infinitesimal generator of the interconnected
system consisting of subsystems (\ref{e.w})~\cite{BC-2009}. 
Consider the vector Lyapunov candidate
for this system, $[V_1(e_1,k)~\ldots V_N(e_N,k)]'$,
with quadratic components $V_i(e_i,k)=e_i'X_i^ke_i$. Also, define 
$V(e,k)=\sum_{i=1}^NV_i(e_i,k)$. 
Since $\mathfrak{L}$ is a linear operator and
$\frac{\partial V_i}{\partial e_j}=0$ for $j\neq i$, then
$[\mathfrak{L}V](e,k)=\sum_{i=1}^m[\mathfrak{L}_iV_i](e,k)$,
where 
\begin{eqnarray*}
\lefteqn{[\mathfrak{L}_i V_i](e,k)\triangleq
\sum\nolimits_{l=1}^M\lambda_{kl}V_i(e_i,l)+\left(\frac{\partial V_i}{\partial
    e_i}\right)^T \Big((A-L_i^kC_i^k)e_i} && \\
&& + \sum\nolimits_{j\in
    \mathbf{V}_i^{k_i}} K_{ij}^k(H_{ij}(e_j-e_i)-G_{ij}w_{ij})\\
&&+(\hat B_2-L_i^k\hat D_i^k)\hat \xi_i 
-\omega_i-\sum\nolimits_{j\in
  \mathbf{V}_i^{k_i}}(\omega_{ij}^{(1)}+\omega_{ij}^{(2)})\Big). 
\end{eqnarray*}

For arbitrary $\tau_i^k,\theta_{ij}^k,\vartheta_{ij}^k>0$, consider the expression
\begin{eqnarray*} 
\lefteqn{[\mathfrak{L}V](e,k)
+ \sum\nolimits_{i=1}^N\Big[
  \tau_i^k(\alpha_i^2 \|C_i^k e_i+\hat D_i^k\hat
  \xi_i\|^2-\|\omega_i\|^2)} && \nonumber \\
&& 
+\sum\nolimits_{j\in\mathbf{V}_i^{k_i}}\theta_{ij}^k(\beta_{ij}^2\|H_{ij}e_i+G_{ij}w_{ij}\|^2-\|\omega_{ij}^{(1)}\|^2)\nonumber \\
&&+\sum\nolimits_{j\in\mathbf{V}_i^{k_i}}\vartheta_{ij}^k(\beta_{ij}^2\|H_{ij}e_j\|^2-\|\omega_{ij}^{(2)}\|^2)\Big]
=
\sum\nolimits_{i=1}^N\mathfrak{R}_i(e,k), 
\end{eqnarray*}
where we let
\begin{eqnarray}
\lefteqn{\mathfrak{R}_i(e,k)\triangleq [\mathfrak{L}_iV_i](e,k) 
  + \tau_i^k\left(\alpha_i^2 \|C_i^k e_i+\hat D_i^k\hat
  \xi_i\|^2-\|\omega_i\|^2\right)} && \nonumber \\
&& 
+\sum_{j\in\mathbf{V}_i^{k_i}}\theta_{ij}^k\left(\beta_{ij}^2\|H_{ij}e_i+G_{ij}w_{ij}\|^2-\|\omega_{ij}^{(1)}\|^2\right)\nonumber \\
&& 
+e_i'\Big(\sum_{j:~
    i\in\mathbf{V}_j^{k_j}}\vartheta_{ji}^k\beta_{ji}^2H_{ji}'H_{ji}\Big)e_i-
\sum_{j\in\mathbf{V}_i^{k_i}}\vartheta_{ij}^k\|\omega_{ij}^{(2)}\|^2. \qquad 
\label{LV.1}
\end{eqnarray}
By completing the squares, one can establish that  
\begin{eqnarray}
\mathfrak{R}_i(e,k)&\le& e_i'U_ie_i+2e_i'X_i^k\sum_{j\in
  \mathbf{V}_i^{k_i}}K_{ij}^kH_{ij}e_j \nonumber \\
&& +\gamma^2(\|\xi\|^2+\|\xi_i\|^2)+ \gamma^2\sum_{j\in
  \mathbf{V}_i^{k_i}}\|w_{ij}\|^2, \label{Rem.2}
\end{eqnarray}
where 
\begin{eqnarray*}
\lefteqn{U_i=X_i^k\left(A-\hat B_2(\hat D_i^k)'(E_i^k)^{-1}C_i^k\right)} &&\\
&&+\left(A-\hat B_2(\hat D_i^k)'(E_i^k)^{-1}C_i^k\right)'
X_i^k+\sum_{l=1}^M\lambda_{kl}X_i^l 
\nonumber \\
&&+\Big(\frac{1}{\tau_i^k}+\!\!\sum_{j\in\mathbf{V}_i^{k_i}}\!\!\Big(\frac{1}{\theta_{ij}^k}+\frac{1}{\theta_{ij}^k}\Big)\Big)X_i^kX_i^k+\!\!\sum_{j:~
    i\in\mathbf{V}_j^{k_j}}\!\!\vartheta_{ji}^k\beta_{ji}^2H_{ji}'H_{ji} 
\\
&&+\frac{1}{\gamma^2}X_i^k
\hat B_2\left(I-(\hat D_i^k)'(E_i^k)^{-1} \hat
    D_i^k\right)\hat B_2'X_i^k 
 \nonumber \\
&&-\gamma^2(C_i^k)'(E_i^k)^{-1}C_i^k
-\gamma^2\sum_{j\in\mathbf{V}_i^{k_i}}H_{ij}'F_{ij}^{-1}H_{ij}.
  \label{Ui}
\end{eqnarray*}
We now observe that it follows from the LMI (\ref{T4.LMI.1}) that for any
nonzero collection of vectors
$e_i,e_j\in \mathbf{R}^n$
\begin{eqnarray}
e_i'U_ie_i+2e_i'X_i^k\sum_{j\in
    \mathbf{V}_i^{k_i}}K_{ij}^kH_{ij}e_j+(p_i^k+q_i^k)\|e_i\|^2 \nonumber \\
-2e_i'\sum_{j\in\mathbf{V}_i^{k_i}}e_j
< \sum_{j=1}^N \pi_{ij}^k (e_j'X_j^ke_j),
\label{Ui.2}
\end{eqnarray}
where  $\pi_{ij}^k$ are elements of the $N\times N$ 
matrix $\Pi^k$, defined as
\begin{equation}
  \label{pi}
  \pi_{ij}^k=\begin{cases} -2\delta_i, & j=i, \\
 \frac{2\delta_j}{q_j^k+1}, & j\in \mathbf{V}_i^{k_i}, \\
0, &\text{otherwise}.
\end{cases}  
\end{equation}
Together with (\ref{Rem.2}), the latter inequality leads to 
\begin{eqnarray}
\lefteqn{\mathfrak{R}_i(e,k)+(p_i^k+q_i^k) \|e_i\|^2-2e_i' \sum_{j\in\mathbf{V}_i^{k_i}} e_j< \gamma^2\|\hat \xi_i\|^2} && \nonumber \\
&& +\gamma^2\sum_{j\in
  \mathbf{V}_i^{k_i}} \|w_{ij}\|^2 +
\sum_{j\in\mathbf{V}_i^k\cup \{i\}} \pi_{ij}^k V_j(e_j,k). 
\label{***}
\end{eqnarray}

It is easy to verify using (\ref{pi}) that all components of the vector
$\mathbf{1}_N'\Pi^k$ are negative and do not exceed $-\epsilon$, where
$\epsilon=\min_{i,k}\frac{2\delta_i}{q_i^k+1}$.
Hence, it follows from (\ref{LV.1}), (\ref{***}) that the following
dissipation inequality holds for all $e_i$, $\xi$, $\xi_i$, 
$w_{ij}$, and 
 for all uncertainty signals $\omega_{i}(t)$,
$\omega_{ij}^{(1)}(t)$, 
$\omega_{ij}^{(2)}(t)$ satisfying the constraints (\ref{wv.constr})      
\begin{eqnarray}
\lefteqn{N\Psi^k(e)+[\mathfrak{L} V](e,k)\le  -\epsilon 
V(e,k) } && \nonumber \\
 && + \gamma^2\sum_{i=1}^N\left (\|\xi_i\|^2+ \|\xi\|^2+ \sum_{j\in
  \mathbf{V}_i^{k_i}} \|w_{ij}\|^2 \right).
\label{lyap.3}
\end{eqnarray}
The statement of Theorem~\ref{T.aux} now follows from (\ref{lyap.3}). This
can be shown 
using the same argument as that used to derive the statement of Theorem~1
in~\cite{U6} from a similar dissipation inequality. 
Indeed, let $\xi,{\xi_i}\in L_2[0,\infty)$, $i=1,\ldots,N$. 
Since equation (\ref{e.w}) defines $(e(t),\eta(t))$ to be a 
Markov process, we obtain from (\ref{lyap.3})  using the Dynkin
formula that 
\begin{eqnarray}
\lefteqn{\mathsf{E}\left[V(e(t),\eta(t))\Big|e(s),\eta(s)\right]-V(e(s),\eta(s))}
&& \nonumber \\
&&
+\mathsf{E}\left[\int_s^t(\epsilon V(e(t),\eta(t))+N\Psi^{\eta(t)}(e(t))dt\Big|e(s),\eta(s)\right]
\nonumber \\
&&\le \gamma^2\mathsf{E}\left[\int_s^t\left(\sum_{i=1}^N\|\xi_i(t)\|^2+
    \|\xi(t)\|^2 \right.\right. \nonumber \\
&&+ \sum_{j=1}^N \mathbf{a}_{ij}^{\eta(t)}\|w_{ij}(t)\|^2 \Bigg)dt\Bigg|e(s),\eta(s)\Bigg].
\label{lyap.5}
\end{eqnarray}
Here $\mathsf{E}\left[\cdot|e(s),\eta(s)\right]$ is the expectation
conditioned on the $\sigma$-field generated by $(e(t),\eta(t)), t\le s$.
We now observe that the processes $ v(t)\triangleq V(x(t),\eta(t))$, 
\begin{eqnarray*}
 \phi(t)&\triangleq& \sum_{i=1}^N\left(\|\xi_i(t)\|^2+
  \|\xi(t)\|^2+\sum_{j=1}^N \mathbf{a}_{ij}^{\eta(t)}\|w_{ij}(t)\|^2\right), \\
  \psi(t)&\triangleq& N\Psi^{\eta(t)}(e(t))+\epsilon V(e(t),\eta(t))
\end{eqnarray*}
satisfy the conditions of Lemma~\ref{Supermartingale.Lemma}. 
This leads to the
conclusion that 
$\int_0^\infty(N\Psi^{\eta(t)}(e(t)))+\varepsilon V(e(t),\eta(t)))dt <\infty$
a.s., and also $\lim_{t\to 
  \infty}V(e(t),\eta(t)) <\infty$ a.s.. Due to the condition $X_i>0$
for all $i$, we conclude that  $\lim_{t\to \infty}
\|e_i(t)\|^2$ exists and $\int_0^\infty\|e_i(t)\|^2dt <\infty$ a.s.. This
implies $\lim_{t\to \infty} e_i(t)=0$ with probability 1 for all $i$ and
arbitrary disturbances $\xi,\xi_i,w_{ij}\in L_2[0,\infty)$; i.e.,
(\ref{convergence.P1}) holds. 

In the case where $\xi_i=0$, $w_{ij}=0$, $\xi=0$, the above observation
immediately yields the statement of the theorem about internal stability of
the system (\ref{e.w}), (\ref{Lim}) with probability
1. The claim of internal exponential mean-square stability follows directly
from (\ref{lyap.3}), since $\Psi^k\ge 0$ by definition.   

Also, by taking the
expectation conditioned on $e_i(0)=x_0$, $\eta(0)=m_0$ on both sides of
(\ref{lyap.5}) and then letting $t\to\infty$, we obtain 
condition (\ref{objective.i.1}), in which $P= \frac{1}{N\gamma^2}
\sum_{i=1}^NX_i^{m_0}$.   
Condition (\ref{convergence}) follows from (\ref{lyap.5}) in a similar
manner. Taking the
expectation conditioned on $e_i(0)=x_0$, $\eta(0)=m_0$ on both sides of
(\ref{lyap.5}), then dropping the nonnegative term $\int_0^tN\Psi^{\eta(t)}dt$ and
letting $t\to\infty$, we establish that
$\mathsf{E}^{x_0,m_0}\int_0^\infty V(e(t),\eta(t)) dt<\infty$. 
Hence $\mathsf{E}^{x_0,m_0}\int_0^\infty\|e(t)\|^2dt <\infty$. 
 \hfill$\Box$

\newcommand{\noopsort}[1]{} \newcommand{\printfirst}[2]{#1}
  \newcommand{\singleletter}[1]{#1} \newcommand{\switchargs}[2]{#2#1}

\end{document}